\documentclass[oneside,letterpaper,10pd]{article}
\pdfoutput=1

\usepackage[letterpaper,total={6.5in, 9in},top=1in]{geometry}

\title{SeMA: A Design Methodology for Building Secure Android Apps} 

\author{Joydeep Mitra \hspace{1cm} Venkatesh-Prasad Ranganath\\
Kansas State University, U.S.A\\ 
\{joydeep,rvprasad\}@k-state.edu
}

\usepackage{float}
\usepackage{verbatim}
\usepackage{listings}
\usepackage{hyperref}
\hypersetup{
    colorlinks=true,
    linkcolor=blue,
    filecolor=magenta,      
    urlcolor=cyan,
}
\usepackage[inline,shortlabels]{enumitem}
\usepackage[plain]{fancyref}
\usepackage{graphicx}

\date{Created: February 15, 2019.  Revised: September 19, 2019.}

\begin{document}

\maketitle

\newcommand{\eg}{e.g.,\,}
\newcommand{\ie}{i.e.,\,}
\newcommand{\etal}{et al.\,}
\newcommand{\app}{\textcolor{red}{application}}
\newcommand{\screen}{\textcolor{red}{screen}}
\newcommand{\tv}{\textcolor{red}{TextView}}
\newcommand{\btn}{\textcolor{red}{Button}}
\newcommand{\init}{\textcolor{red}{init}}
\newcommand{\trans}[2]{\textcolor{red}{go\hspace{1mm}from}\hspace{1mm}#1\hspace{1mm}{\textcolor{red}{to}}\hspace{1mm}#2}
\newcommand{\cond}{\textcolor{red}{condition}}
\newcommand{\prop}[2]{\textcolor{red}{propagate}\hspace{1mm}#1\hspace{1mm}\textcolor{red}{as}\hspace{1mm}#2}
\newcommand{\pressed}{\textcolor{red}{was\hspace{1mm}pressed}}
\newcommand{\MyAnd}{\textcolor{red}{and}}
\newcommand{\MyOr}{\textcolor{red}{or}}
\newcommand{\MyNot}{\textcolor{red}{not}}
\newcommand{\res}{\textcolor{red}{resources}}
\newcommand{\Sreq}{\textcolor{red}{security-requirements}}
\newcommand{\private}{\textcolor{red}{is\hspace{1mm}private}}
\newcommand{\jm}[1]{\textcolor{red}{#1}}

\begin{abstract}
    UX (user experience) designers visually capture the UX of an app via storyboards. This method is also used in Android app development to conceptualize and design apps.

    Recently, security has become an integral part of Android app UX because mobile apps are used to perform critical activities such as banking, communication, and health. Therefore, securing user information is imperative in mobile apps.

    In this context, storyboarding tools offer limited capabilities to capture and reason about the security requirements of an app. Consequently, security cannot be baked into the app at design time. Hence, vulnerabilities stemming from design flaws can often occur in apps. To address this concern, in this paper, we propose a storyboard based design methodology to enable the specification and verification of security properties of an Android app at design time.
\end{abstract}

\section{Why a new methodology?}
\label{sec:motiv}

Android app development teams use storyboarding as part of their design process \cite{Marcus:2013, DesTech:URL}. A storyboard is a sequence of images that serves as a specification of the user observed behavior in terms of screens and transitions between screens. Storyboarding helps designers identify different kinds of app users (user profiles), explore possible real world scenarios in which a user will interact with the app, and develop wireframes to capture the scenarios for identified user profiles \cite{DesTech:URL,UXMag:URL,Iedf:URL}. Numerous tools such as Xcode \cite{Xcode:URL}, Sketch \cite{Sketch:URL}, and Android JetPack's Navigation Component \cite{JetPackNav:URL} help designers digitally express the storyboard of an app.



Storyboarding is meant to be participatory because it can be used by designers to get feedback from potential users about likely user scenarios, and from developers about the technical challenges of implementing the design. However, in their current form, storyboards do not capture non-functional requirements such as security.  Consequently, Android app designers and developers cannot use storyboarding to collaborate and express security requirements at design time.  As a result, verification of security requirements is delayed until later stages.  This delay increases the cost of app development as vulnerabilities due to flaws detected in later stages increase the cost of development \cite{Telang:2007}.

In this context, to enable reasoning about security at design time, we are exploring two ideas: \emph{1) extend storyboards to capture non-functional properties and 2) develop a methodology that uses storyboards to specify and verify security requirements of apps at design time.}

To understand the reasons for and benefits of these ideas, we need to understand the current landscape of Android app development practices.

Today, developers consider the security requirements of an app while or after implementing the app. Existing research efforts related to Android app security have focused on developing tools and techniques to help detect vulnerabilities in an app's implementation \cite{Sufatrio:2015,Li:2017}. Despite such efforts, apps with known vulnerabilities find their way to app stores \cite{Sadeghi:FASE14,document:gomez14}. This is possibly because existing tools are neither accurate \cite{Ranganath:TR19} nor scalable \cite{Pauck:2018} in terms of detecting known vulnerabilities.
Given this landscape, we believe it is worthwhile to explore the unexplored: \textit{preemptively eliminate vulnerabilities in apps by enabling developers and designers to consider security at design time and thwart the exploits of malicious apps}.

Besides complementing existing techniques, such exploration will help consider the following research questions:

\begin{enumerate}[label=\alph*)]
    \item How much effort is required in terms of time and cost to formally reason about Android apps at design time?
    \item In the context of Android app development, is secure-by-design cost-effective?
    \item Is verifying an app based on its storyboard easier than verifying an app based on its code?
    \item In terms of improving security, how will the proposed approach compare to existing curative approaches, \ie detect vulnerabilities after they occur?
\end{enumerate}

\section{What is the methodology?}
\label{sec:proposal-1}


Motivated by the above reasons, we propose the following methodology, \textit{Securing Mobile Apps (SeMA)}, that borrows heavily from model driven development \cite{France:2007, Hatcliff:2003}.
\begin{enumerate}[label=\arabic*)]
  \item App development begins with the storyboard (model) of an app that captures the screens of the app, the transitions between screens, and the resources used by the app.
  \item The storyboard is iteratively refined by adding behavioral and security properties.
  \item As the storyboard is refined, verification techniques are used to check the behavior of the storyboard satisfies the security properties.
  Once the refined storyboard satisfies the desired requirements, property preserving structural code\footnote{Structural code is traditional skeletal code along with some non-trivial logic to enforce properties.} of the app is automatically generated.
  \item Business logic is added to the generated code.
\end{enumerate}


\subsection{An Illustrative Example}

To understand SeMA, consider the development of an app that sends a message to a set of pre-identified contacts at the push of a button.

\subsubsection{App Specification}
\label{sec:appSpec}

\begin{figure*}[t]
    \centering
    \includegraphics[height=13cm]{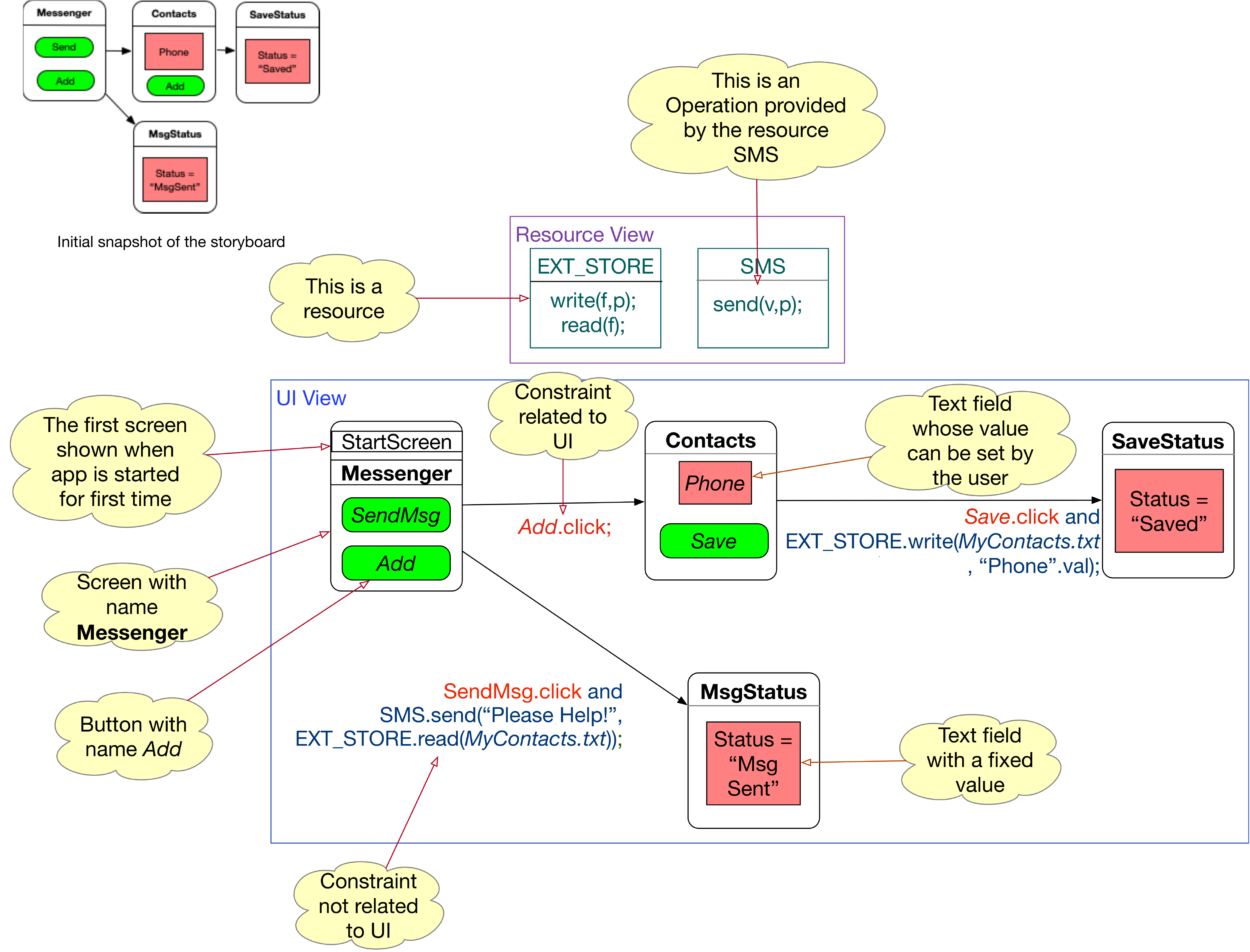}
    \caption{Storyboard of the Emergency App. The clouds are not part of the storyboard; they provide information about the different graphical elements in the figure.}
    \label{fig:ex-1}
\end{figure*}

\Fref{fig:ex-1} shows the storyboard of this example app along with a brief description of various concepts captured in the storyboard.

Initially, the app designer specifies the required screens and the transitions between the screens as shown in the sub-diagram in the top left corner of the figure (labeled A). The initial snapshot is similar to how storyboards are used in designing apps today.  Besides the graphical description of screens of the app, this snapshot captures the navigational possibilities in the app: 1) user will see the \textit{Messenger} screen up on launching the app, 2)  the user can transition to either \textit{Contacts} or \textit{MsgStatus} screens from the \textit{Messenger} screen, and 3) the user can transition to \textit{SaveStatus} screen from the \textit{Contacts} screen.

Next, the app designer refines the initial snapshot with constraints based on user actions (marked in red in the main diagram). For example, when the user clicks the \textit{Add} button in the \textit{Messenger} screen, the app transitions to the \textit{Contacts} screen.

As the next refinement, the app developer (in collaboration with the app designer)adds more behavioral constraints to the storyboard.  For example, a constraint based on the state of the app (marked in blue) is added to the transition from the \textit{Contacts} screen.  Other transitions in the app are enriched with constraints in subsequent refinements.

Many constraints rely on operations involving the resources in the device. While such resources are pre-defined, apps can specify views of resources that they will use.

The \textit{resource view} of the app (shown in \Fref{fig:ex-1}) describes the view of the resources used in the example app as follows.

\begin{itemize}
  \item \textit{EXT\_STORE} denotes the external storage on Android devices that can be accessed by all apps installed on the device. The app uses the following two capabilities of this resource: \texttt{write(f,p)} writes the value \texttt{p} into file \texttt{f} and \texttt{read(f)} reads a value from file \texttt{f} and returns it.
  \item \textit{SMS} denotes a pre-defined service available on Android devices that the app uses to send a SMS message \texttt{m} to a list of phone number \texttt{p} via \texttt{send(m,p)} operation.
\end{itemize}

The final storyboard with all the behavior and resource operations specified will look like \Fref{fig:ex-1}.

To give an example of how the constraints in a transition are interpreted, the transition from the \textit{Messenger} screen to the \textit{MsgStatus} screen is taken when 1) the \textit{SendMsg} button is pressed and 2) an SMS message is successfully sent to the phone numbers in the file named \textit{MyContacts.txt}.

\subsubsection{Analysis}
\label{sec:analysis}

In this example, suppose we are interested in the violation of this property: \textit{malicious input should not influence data flow in the app.}




Based on the semantics of \textit{EXT\_STORE}, all apps can access data in \textit{EXT\_STORE}.  Hence, \textit{EXT\_STORE} can possibly contain malicious inputs as malicious apps can modify the contacts stored by the example app. Since the phone numbers from \textit{EXT\_STORE} is used in \textit{SMS.send(\ldots)}, a violation of the property is detected by analyzing the data flow through the storyboard in accordance with the semantics of resources used by the app.

This vulnerability can be fixed by storing the phone numbers in a more secure way. For example, \textit{MyContacts.txt} file can be stored in the app's internal storage that is accessible only by the app.  This fix can be suggested by tooling support, expressed by the developer in the storyboard by using appropriate resources, and enforced via code generation.

\section{Realizing the methodology}
To validate the feasibility of this methodology, we are currently developing the design language, its formal syntax and semantics, and the analysis engine required to reason about the design.  In addition, we are exploring the following few design choices.

\subsection{Leverage existing mobile app design frameworks}
Android JetPack Navigation (AJN) \cite{JetPackNav:URL} is a suite of libraries that helps Android developers design their apps' navigation. Navigation
refers to the interactions that allow users to move between screens with simple button clicks and other more complex patterns.
Since AJN is open source, is already supported in Android Studio, and can easily be extended, we chose to extend AJN with functionality to enable the specification and analysis of an app's design as described in Section \ref{sec:appSpec}. The modified library currently allows users to do the following:

\begin{enumerate}
  \item Initialize the widgets of a screen with data.
  \item Annotate transitions between screens with constraints.
  \item Pass data as arguments on a transition into a screen.
\end{enumerate}

\subsection{Support pre-defined security properties}
We do not want to burden app developers or designers with specifying the security properties required for verification because specifying such properties may initially be a non-trivial burden for the average app developer or designer. Moreover, requiring app developers or designers to specify such properties will unnecessarily complicate the existing design process.
Therefore, we envision SeMA will initially support the checking of pre-defined security properties that are based on classes of common vulnerabilities \cite{Mitra:2017} \eg listed in \Fref{sec:future}.

\subsection{Leverage existing tooling framework}
Android Lint \cite{Lint:URL} is a static code analysis tool that helps analyze Android project source files. We have extended Android Lint with
customized checks on navigation graphs. Currently, our extension to Android Lint can check the following properties:

\begin{enumerate}
  \item Data accepted as input via an exported screen\footnote{An exported screen can be triggered by any component outside the app.} is used to perform a sensitive operation in the app.
  \item Data from an untrusted source (\eg app's external storage) is used to perform a sensitive operation in the apps.
\end{enumerate}

To check these properties, we rely on Android Lint to identify data sources and sinks.  Then, our Lint extension tracks the data flow between data sources and data sinks.  A violation is flagged when data flows from a potential malicious data source into a data sink associated with sensitive operations.

We use the JavaPoet library \cite{JavaPoet:URL} to translate the verified navigation graph into Android source code. The generated source code can be extended by developers with business logic. Such extensions might introduce behavioral inconsistencies between the storyboard and the implementation.  To detect such inconsistencies, we plan to generate characterization tests based on the behavioral specifications captured in the storyboard.

Further, not all security properties can be checked statically at the design stage.  Such properties can be checked preemptively at runtime, and we plan to explore this possibility.

\subsection{Stitching a Workflow}
As described in \Fref{sec:proposal-1}, the methodology will help the user build an app iteratively. First, the user uses the extended version of AJN to create an app's storyboard.  Second, this storyboard is refined with behavioral properties. Third, the storyboard is verified against pre-defined properties after every refinement.  Fourth, the structural code is generated once the refined storyboard satisfies the desired properties. Finally, the user extends the structural code to build a working app.  The extended code is then tested against the generated characterization tests to ensure that the implementation is consistent with the design.  Currently, we have prototyped the methodology (with the exception of characterization test generation step) in Android Studio.

\section{What lies in the future?}
\label{sec:future}
Going forward, we plan to extend AJN with more features that will enable the analysis of navigation graphs.  We will also extend the custom lint checker with more properties based on known vulnerabilities that plague Android apps.

Existing efforts have identified vulnerabilities that commonly plague Android apps \cite{Chin:Thesis13} and have developed benchmarks to capture them \cite{Mitra:2017}.
Most known vulnerabilities can be classified as follows:

\begin{enumerate}
    \item{\textit{Reliance on Data from Potentially Malicious Sources}}:
    If an app X blindly trusts input from an external component not in the app and uses it to manipulate data or perform a critical task, then the app is vulnerable.

    \item{\textit{Disclosure of Sensitive Information}}:
    An app can accidentally expose sensitive information.

    \item{\textit{Exposure of Privileged Resources}}:
    An app with access to privileged resources can accidentally allow other apps to access to such resources and enable \textit{privilege escalation}.


\end{enumerate}

The above classes of vulnerabilities can be prevented by SeMA.

As for vulnerabilities that stem from implementation errors (e.g. using HTTP instead of HTTPS when contacting a server).
They can be prevented by smart code generation techniques.

\section{Challenges}
\label{sec:chal}

To successfully realize the proposed methodology, the following interesting challenges need to be addressed while developing SeMA.

\begin{enumerate}

\item \textit{Storyboard Extensions to Capture Non-UI Behavior}: Today, storyboards of Android apps capture only the UI components of an app and behavior triggered by user actions. However, Android apps also exhibit non-UI related behavior, \eg reading a file from the app's storage as shown in the earlier illustrative example. Therefore, existing storyboards will have to be extended with features to capture behavior that stem from non-UI elements without disrupting the workflow and the mental model associated with storyboards. 

\item \textit{Context-Aware Analysis}: Android apps do not run in isolation. They communicate with remote servers, other apps on the device, and the underlying Android framework. Therefore, the security of an app depends on the context in which it is operating. Security analysis of app storyboards will have to be aware of such contexts; otherwise, the analysis will be imprecise.

    \item \textit{Reactive Nature of Android Apps}: Android apps interact with the underlying platform via platform-defined application lifecycle methods to handle the numerous system and user events. These events could trigger the execution of an app from different entry points.  Also, apps have the ability to persist information across different executions.  This allows for different executions of an app that start at different entry points to be implicitly related by persisted data. This possibility should be considered to accurately reconstruct the behavior of an app for the purpose of security reasoning.

    \item \textit{Scalability of Analysis}: To be accurate and effective, security analysis of Android apps will have to consider various contexts and rich features of the Java (or Kotlin) programming language used to implement the apps.  Such analysis is neither scalable \cite{Pauck:2018} nor accurate \cite{Ranganath:TR19} as they are plagued by the challenges faced by non-trivial source code analysis \cite{Li:2017}.

    \item \textit{Checking Richer Security Properties}: Given the limitations of simple static checkers like Android Lint and the richness of security properties (\eg temporality), more involved analysis/checking techniques may be needed to check/enforce richer security properties.  Such techniques could be adopted from program analysis/verification community or custom techniques could be developed from scratch.

    \end{enumerate}

\bibliography{references.bib}
\bibliographystyle{plain}

\end{document}